\documentclass[11pt]{article}
\usepackage{moriond,epsfig}

\def\be{\begin{equation}}
\def\ee{\end{equation}}
\def\bea{\begin{eqnarray}}
\def\eea{\end{eqnarray}}

\begin{document}
\vspace*{3cm}
\title{HARP: a hadron production experiment}

\author{ Emilio Radicioni }

\address{CERN (Geneva, Switzerland) and INFN-Bari (Bari, Italy) \\ 
         on behalf of the HARP collaboration}

\maketitle\abstracts{
Hadron production is a key ingredient in many aspects of neutrino physics.
Precise prediction of atmospheric neutrino fluxes, characterization of 
accelerator neutrino beams, quantification of pion production and capture 
for neutrino factory designs, all of these would profit of high-statistics, 
high-precision hadron production measurements.
In addition, such data is needed for the calibration of Monte-Carlo hadron 
production models.
The HARP experiment at CERN is performing extensive measurements of 
cross sections and secondary particle yields to address these problems.
}

\section{Introduction}

Hadron production data is relevant in several branches of neutrino physics.

The study of atmospheric neutrinos provides strong evidence for neutrino 
oscillations. At the same time, to quantitatively understand this phenomenon, 
several accelerator-based neutrino  experiments are being built, and new types
of neutrino beams are being designed.
In either cases, detailed knowledge of the hadron cross sections at the
relevant energies is now considered a must.

The design of alternative neutrino beams would profit a lot from a 
more detailed knowledge of the hadron production cross-sections. Optimizing 
the figure of merit (and the cost) of the so-called neutrino factory (neutrino 
beam produced by decays in a muon storage ring) depends on the production
and collection efficiency of pions at the target station, and therefore
requires a systematic study of pion production cross sections at 
several energies (from a few GeV to 24 GeV, depending on design choices) 
and with several target materials.

The calculation of atmospheric neutrino fluxes is dominated by the primary
cosmic ray flux, by the hadron production in the interaction of the primaries
with target air nuclei, and by the subsequent decay chains.
Since the primary cosmic ray spectrum is considered to be known to better than
10\%, most of the uncertainty comes from the limited understanding of hadron
interactions. Different Monte-Carlo simulations, depending on the model they are
based on, provide estimates which can differ by as much as 30\% \cite{Engel:1999zq}.
The relevant energy range for primary particles is, in this case, from a few GeV 
to a few 100 GeV, while target material should be as close as possible to the 
constituents of the atmosphere, namely $N_2$ and $O_2$.

Traditional neutrino beams are produced, much similarly as the atmospheric 
neutrinos, as decay products of the particles produced by a primary interaction
of a proton beam in a target. It is therefore not surprising that, again, 
being able to describe the beam composition in detail relies on the 
detailed knowledge of the primary hadron interaction cross sections. 
The range of interesting beam energies spans from 8GeV \cite{Church:1997ry} (for the 
beam of the MiniBooNE experiment at FNAL), to 12GeV (K2K beam \cite{Oyama:1998bd} from KEK to
SuperKamiokande), to 50GeV (the future JHFnu beam, again to SuperKamiokande)
to 120GeV (to MINOS from FNAL), to 400GeV (CERN's CNGS).

A valuable by-product of such a systematic study would be a calibration sample
for Monte-Carlo hadron production models, very much useful not only in
neutrino physics, but also as precious a tool in the study of present and future
detectors performances.

From this point of view, it is not surprising that relatively
low-energy (in the 1GeV-to-100GeV scale) hadron production experiments are being
operated, built or proposed  ~\cite{Catanesi:1999xr} ~\cite{Catanesi:2001gi} ~\cite{Chemakin:1999cp} ~\cite{p907}. Such experiments all share a basic design, consisting
in the presence of open-geometry spectrometers, as close as possible to full
angular coverage, and in the aim at full particle identification.
This design has already been seen in heavy-ions experiments, whose calibration
samples with proton beams are similar to the kind of measurements needed for 
our purposes. However, they are limited in acceptance or, most frequently, in statistics 
(small event sample, only one beam momentum, limited number of target materials).

\begin{figure}
	\center \mbox{\epsfxsize=11cm \epsffile{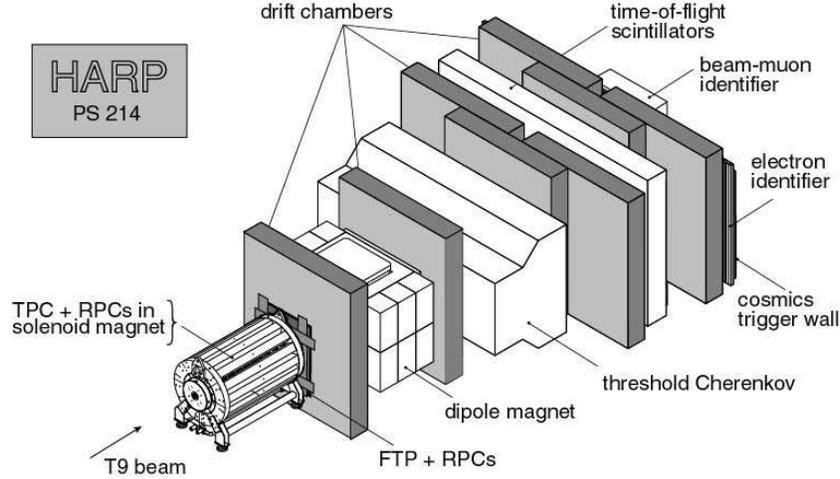}}
	\caption{ The HARP apparatus.}
	\label{apparatus}
\end{figure}

The HARP \cite{Catanesi:1999xr} experiment at CERN is performing a systematic study of hadron 
cross sections at several beam momenta (from 2GeV/c to 15GeV/c) and on a large
number of thin, thick and cryogenic targets. HARP's design (Fig. \ref{apparatus})
aims at a full-acceptance 
apparatus, capable of collecting a large number of settings thanks to its high-rate 
capability of one {\em setting}, corresponding to a few million events, per day. The
desired overall precision of about 2\% requires keeping the efficiency under
control down to the level of 1\%, primarily by redundancies in the acceptance
regions of the detectors, shown in figure \ref{pid}.

In addition to the standard set of target geometries and materials, 
HARP is also measuring production cross section on special targets provided 
by the K2K and MiniBooNE collaborations.

\section{HARP apparatus}


The HARP experiment is located on the T9 beam line in the East Hall of
the CERN PS complex. 

The PS delivers 24GeV/c protons to a primary target, 
supplying secondary beams to the East Hall experimental area. Several 
bending and focusing magnets allow to momentum-select and steer
the beam to a 10mm spot onto the HARP target. The PS is operated with a 
16.8s super-cycle; during this time, it delivers several spills to 
the East Hall in slow-extraction mode with a 400ms flat-top. HARP has
been designed with the aim of collection up to 1000 events in such
spills, which represents a relatively high rate, especially for the TPC.
Since the beam is 
an admixture of several particle types (deuterons, protons, kaons, pions, 
muons, electrons) all of the same momentum, the beam line is instrumented
with threshold Cerenkov and Time-Of-Flight detectors to tag the incoming
particle.
To precisely predict the impact position on the HARP target, trajectories 
of beam particles are recorded by 3 sets of X-Y multi-wire proportional 
chambers.

Particles hitting the target are defined by a 2cm diameter scintillation
counter suspended on the beam line, which is then used in coincidence with
a forward scintillator plane and a barrel-shaped scintillating fiber 
trigger surrounding the target.

HARP is using several thin (2\% $\lambda_i$ and 5\% $\lambda_i$) and thick
(0.5 to 1 $\lambda_i$) targets. Available materials are carbon, aluminum,
copper, tin, tantalum and lead.
In addition, cryogenic (liquid $H_2, D_2, N_2, O_2$) and special
{\em experimental} (K2K and MiniBooNE)
targets are also available.
Data points will be collected at several momenta from 2GeV/c to 15GeV/c, 
for both positive and negative beam polarities.


\subsection{Large-angle detectors}

Particles with high transverse momentum are detected in a {\em large-angle spectrometer},
made by the Time Projection Chamber (TPC) for particle identification by tracking 
in magnetic field and  measurement of $dE/dx$, and by a high-precision 
Resistive Plate Chambers (RPCs) time-of-flight arranged in a cylinder around the TPC.
To allow backward-going particles to be detected, the target is placed inside the 
TPC volume.

The TPC consists in a cylindrical volume 1.5m long and 0.8m diameter filled with a 91\% Ar, 
9\% $CH_4$ gas mixture. The trajectories of charged particles are bent by a 0.7T 
solenoid magnetic field.
A 12KV electric field drives the ionization charges at a velocity
of $5 cm/\mu s$ to the read-out plane, where the induction signals are collected by 
3972 pads arranged in 20 concentric rows. The pad signals are digitized in 100ns time bins,
corresponding to about 5mm bins in the longitudinal direction.

The RPC system is made by 4-layers, 0.3mm gap RPC chambers. They allow
for Time-Of-Flight measurement, $e-\pi$ separation in the 100-250 MeV/c region
and rejection of tracks
from piled-up events by measuring the time of crossing tracks with a precision 
of about 150ps.

\subsection{Forward detectors}

Immediately after the TPC, a {\em forward trigger} scintillator plane selects events
with small-angle particles. It is complemented by a further set of RPC chambers for $e-\pi$ 
separation.
Particles leaving the target in the forward direction are poorly seen in 
the TPC, and are therefore measured and tagged by the {\em forward spectrometer}.

The spectrometer is built around a 4-station Drift Chamber system, where a 0.68T dipole 
magnet is inserted between the first and the second station. The chambers are filled with
90\% Ar, 9\% $CO_2$ and 1\% $CH_4$. The drift chamber modules have been designed for
the NOMAD experiment. They are made by 4 sets
of 3 planes inclined at angles of $-5^°$, $0^°$, $+5^°$, and in HARP they have been
oriented to have vertical wires. The drift chambers are expected to deliver $200\mu$ 
resolution in the horizontal plane; this is a lower figure than quoted in Nomad, due to 
a change in the gas mixture imposed by safety constraints.

A $30 m^3$ Cerenkov detector sits after the second drift chamber station. It is filled
with $C_4F_{10}$ at atmospheric pressure. A set of cylindrical mirrors collect the light 
to 38 photo-multipliers. Its main purpose is the particle identification in the high-energy
region. The thresholds are 2.6GeV/c for $\pi$, 9.3GeV/c for K and 17.6GeV/c for p.

\begin{figure}
	\center
	\mbox{\epsfxsize=7.5cm \epsffile{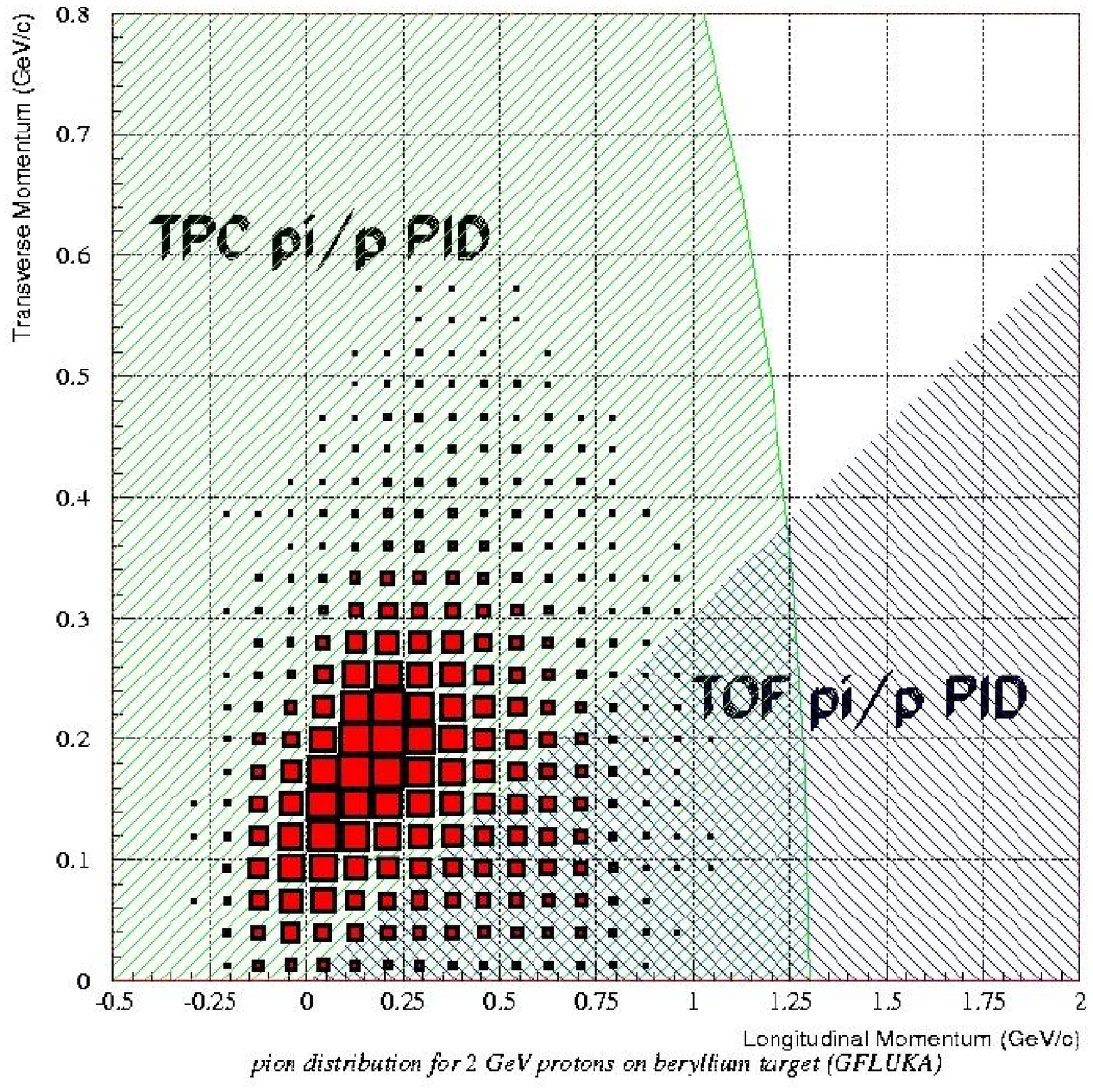}}
	\mbox{\epsfxsize=7.5cm \epsffile{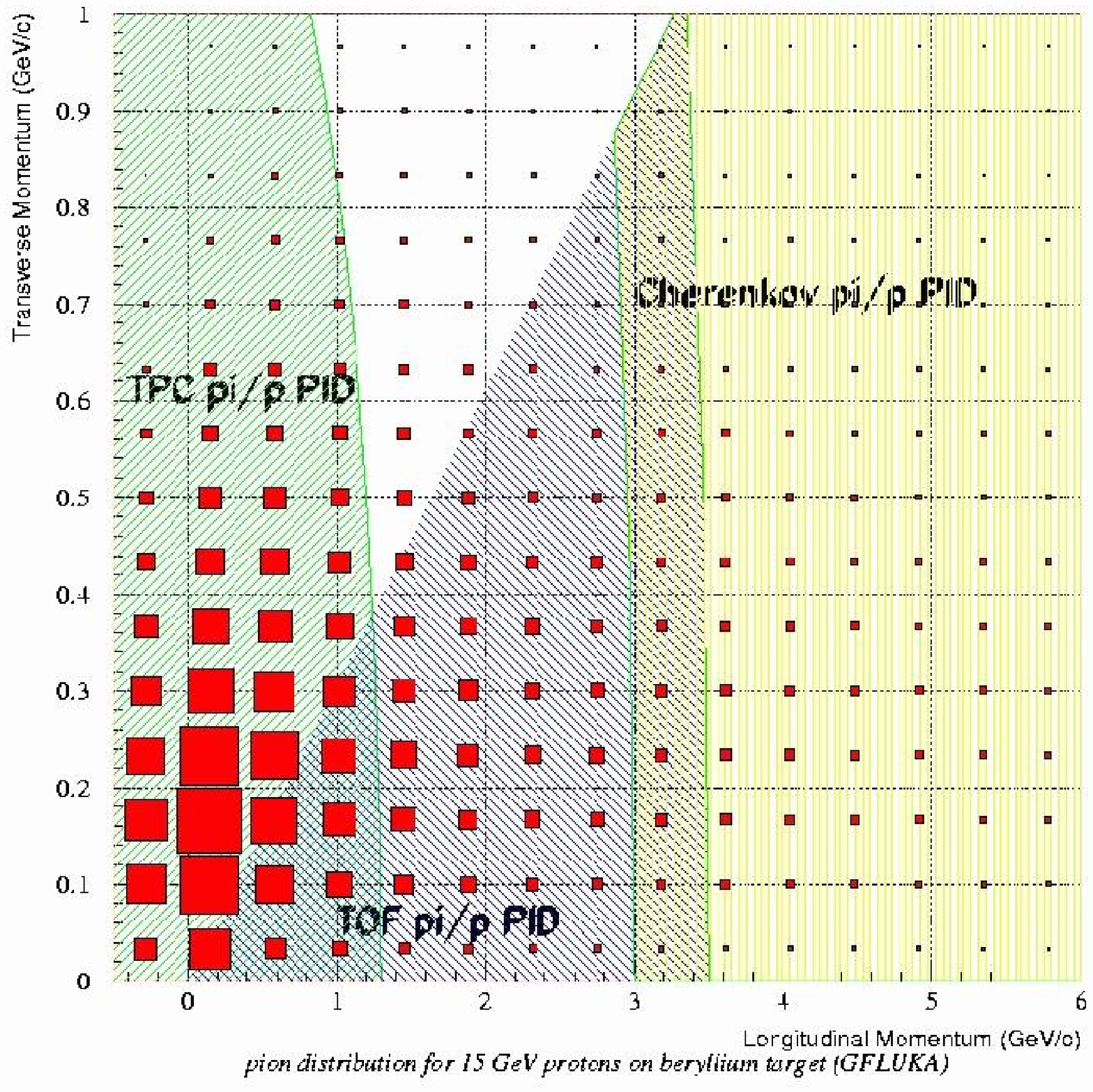}}
	\caption{ Particle Identification.}
	\label{pid}
\end{figure}

After the third drift chamber station, a large Time-Of-Flight wall made of 39 scintillation
counters identifies particles in the intermediate-energy range where, as shown 
in figure \ref{pid}, 
it links the TPC to the Cerenkov. A preliminary analysis of data taken last year
indicates a resolution of about 200ps.

At the far end of the apparatus, a lead/scintillating-fibers calorimeter is used as an
electron identifier. This is a refurbished version of the CHORUS electron/hadron 
calorimeter.

\section{Present status}

After extensive debugging of the detectors during the winter shutdown, data taking has
restarted on May 6th 2002, and will continue until end of September.

In the meantime, first attempts at analyzing the data from the year 2001 
have been concentrated on low-energy points, with the aim of studying the beam 
composition and contamination, the RPC calibration and resolution, and the calibration, 
equalization, clustering, track finding and fitting in the TPC.
Very preliminary momentum distributions have been produced and presented to the CERN-SPSC. 
First published results will be available soon.

\end{document}